\documentclass[%
 aip,
 amsmath,amssymb,
reprint,%
]{revtex4-1}

\usepackage{graphicx}% Include figure files
\usepackage{dcolumn}% Align table columns on decimal point
\usepackage{bm}% bold math
\usepackage[utf8]{inputenc}
\usepackage[T1]{fontenc}
\usepackage{mathptmx}
\usepackage{etoolbox}
\usepackage{color}

\makeatletter
\def\@email#1#2{%
 \endgroup
 \patchcmd{\titleblock@produce}
  {\frontmatter@RRAPformat}
  {\frontmatter@RRAPformat{\produce@RRAP{*#1\href{mailto:#2}{#2}}}\frontmatter@RRAPformat}
  {}{}
}%
\makeatother
\begin{document}

%\preprint{AIP/123-QED}

\title{High-entropy effect at rare-earth site in DyNi}
% Force line breaks with \\
\author{Yuito Nakamura}
\affiliation{Department of Electrical Engineering, Faculty of Engineering, Fukuoka Institute of Technology, 3-30-1 Wajiro-higashi, Higashi-ku, Fukuoka 811-0295, Japan}%Lines break automatically or can be forced with \\

\author{Koshin Takeshita}
\affiliation{Department of Electrical Engineering, Faculty of Engineering, Fukuoka Institute of Technology, 3-30-1 Wajiro-higashi, Higashi-ku, Fukuoka 811-0295, Japan}%Lines break automatically or can be forced with \\

\author{Terukazu Nishizaki}
\affiliation{Department of Electrical Engineering, Faculty of Science and Engineering, Kyushu Sangyo University, 2-3-1 Matsukadai, Higashi-ku, Fukuoka, 813-8503, Japan}

\author{Jiro Kitagawa}
 \email{j-kitagawa@fit.ac.jp}
 \affiliation{Department of Electrical Engineering, Faculty of Engineering, Fukuoka Institute of Technology, 3-30-1 Wajiro-higashi, Higashi-ku, Fukuoka 811-0295, Japan}%Lines break automatically or can be forced with \\

\date{\today}% It is always \today, today,
             %  but any date may be explicitly specified

\begin{abstract}
We report the structural and magnetic properties of RNi (R=Dy, Tb$_{1/3}$Dy$_{1/3}$Ho$_{1/3}$, and Gd$_{1/5}$Tb$_{1/5}$Dy$_{1/5}$Ho$_{1/5}$Er$_{1/5}$) to investigate the high-entropy effect at the rare-earth site.
The lattice parameters are almost unchanged by the increase of configurational entropy, which is due to the successive partial substitution of Dy by pair of rare earth elements located on both sides of Dy in the periodic table.
All compounds exhibit ferromagnetic ground states.
The replacement of Dy with Tb+Ho, which does not have magnetic interactions in competition with Dy, does not affect the magnetic ordering temperature.
Although (Gd$_{1/5}$Tb$_{1/5}$Dy$_{1/5}$Ho$_{1/5}$Er$_{1/5}$)Ni shows the Curie temperature close to that of DyNi, an additional magnetic anomaly, which would be a spin reorientation, is observed probably due to the introduction of competing magnetic interactions between R=Gd and Er compounds and R=Tb, Dy, and Ho ones.
We have also assessed the magnetocaloric effect, and the configurational entropy dependence of the magnetic entropy change reflects that of the temperature derivative of the magnetic susceptibility. 
Our analysis suggests the possibility of enhancing magnetocaloric properties by designing the anisotropy of rare-earth magnetic moments in the high-entropy state.
\end{abstract}

\maketitle

\section{Introduction}
High-entropy alloys (HEAs) are unique systems composed of multiple elements with near equimolar ratios.
They offer a vast compositional space and are a promising platform for studying novel phenomena\cite{Zherebtsov:Int2020,Kitagawa:APLMater2022,Baba:Materials2021}.
Additionally, they have attracted considerable attention due to their rich functionalities, such as high strength, energy storage, radiation protection, magnetism, superconductivity, and biocompatibility\cite{Sathiyamoorthi:PMS2022,Chang:AM2020,Wang:JMCA2021,Marques:EES2021,Pickering:Entropy2021,Chaudhary:MT2021,Kitagawa:JMMM2022,Kitagawa:Metals2020,Castro:Metals2021}.
HEA concept is now introduced into intermetallic compounds (high-entropy intermetallic compounds).

\begin{table*}
\centering
\caption{\label{tab:table1}Lattice parameters ($a$, $b$, and $c$), and magnetic ordering temperature $T_\mathrm{C}$ of each sample.}
\begin{ruledtabular}
\begin{tabular}{ccccc}
 sample & $a$ (\AA) & $b$ (\AA) & $c$ (\AA) & $T_\mathrm{C}$ (K) \\
\hline
DyNi & 7.026(4) & 4.176(2) & 5.444(3) & 59 \\
(Tb$_{1/3}$Dy$_{1/3}$Ho$_{1/3}$)Ni & 7.029(5) & 4.171(3) & 5.439(3) & 63  \\
(Gd$_{1/5}$Tb$_{1/5}$Dy$_{1/5}$Ho$_{1/5}$Er$_{1/5}$)Ni & 7.044(6) & 4.168(3) & 5.447(4) & 63 \\
\end{tabular}
\end{ruledtabular}
\end{table*}

Numerous rare-earth intermetallic compounds exhibit magnetic moments solely attributed to the rare-earth elements.
However, the influence of a high-entropy state at the rare-earth site on the magnetic ordering temperatures of such systems remains insufficiently explored.
We are primarily concerned with the robustness of the magnetic ordering of rare-earth atoms in the presence of the high-entropy state.
In this study, we focus on the well-defined RNi (R:rare-earth) system, wherein magnetic ordering temperatures and magnetic structures are elucidated.
The highest magnetic ordering temperature\cite{Rajivgandhi:JMMM2017} is 71 K in GdNi.
The ordering temperature is moderately lower compared to R$_{2}$In or R$_{6}$CoTe$_{2}$ series, where Gd$_{2}$In and Gd$_{6}$CoTe$_{2}$ show the highest magnetic ordering temperatures of 190 K and 220 K, respectively\cite{Franco:PMS2018}.
Hence, we anticipate the possible destruction of magnetic orderings within all RNi compounds by increasing atomic disorder.

Additionally, we are concerned with the potential modulation of magnetocaloric effects by introducing a high-entropy state. 
Certain RNi compounds demonstrate a significant magnetocaloric effect in proximity to the temperature of liquid hydrogen\cite{Rajivgandhi:JMMM2017,Zhao:JMST2021}.
This observation holds promise for magnetic refrigeration-based hydrogen liquefaction and is significant in realizing a hydrogen society.
The magnetocaloric effects of HEAs have garnered considerable attention\cite{Chaudhary:MT2021,Perrin:JOM2017,Law:APLMater2021,Perrin:Enc2022,Law:JMR2023}. 
Notably, the equimolar quinary alloy GdTbDyHoEr exhibits a remarkable magnetocaloric effect\cite{Yuan:AM2017}. 
A recent investigation into the configurational entropy dependence of magnetocaloric effects in rare-earth HEAs has revealed that magnetic properties depend on the intrinsic magnetic characteristics of rare-earth elements\cite{Lu:JALCOM2021}.
Another study\cite{Law:JMR2023} suggests a reduction in the peak value of magnetic entropy change with an increase in configurational entropy in HEAs containing Dy. 
Transition-metal-based HEAs, such as FeMnNiGeSi, have emerged as a novel material class enabling the manipulation of magnetocaloric effects by introducing magneto-structural transformations\cite{Law:AM2021}.
To the best of our knowledge, reports on the magnetocaloric effects of crystalline high-entropy rare-earth intermetallic compounds are rare, while there are many reports for amorphous HEAs containing rare-earth and transition-metal elements\cite{Law:JMR2023}.

It is well-established that the lattice parameters and the number of 4$f$ electrons significantly impact the magnetic properties in rare-earth intermetallic compounds. 
So, we examined the configurational entropy dependence of the magnetic properties of DyNi through a successive replacement of Dy with a pair of rare-earth elements located on both sides of Dy in the periodic table: partial replacement by Tb+Ho or Gd+Tb+Ho+Er.
Within our replacement sequence, we can maintain the lattice constants and the average number of 4$f$ electrons. 
Consequently, we could explore the high-entropy effect at the rare-earth site while preserving the electronic state of DyNi intact.
In RNi compounds, GdNi and (Dy, Ho, or Er)Ni crystallize into the orthorhombic CrB-type and the orthorhombic FeB-type structure, respectively\cite{Rajivgandhi:JMMM2017,Paudyal:PRB2008,Sato:JMMM1986}.
The crystal structure of TbNi might be controversial: a monoclinic structure with the space group {\it P2$_{1}$m} or an orthorhombic with the space group {\it Pnma}\cite{Rajivgandhi:JMMM2017}.
All RNi (R=Gd to Er) compounds are ferromagnets with the Curie temperature $T_\mathrm{C}$=71 K for R=Gd, 67 K for R=Tb, 62 K for R=Dy, 37 K for R=Ho, and 13 K for R=Er, respectively\cite{Gignoux:JMMM1976,Rajivgandhi:JMMM2017}.
Despite the changes in crystal structure that occur upon going from R=Gd to R=Tb and from R=Tb to R=Dy, we synthesized DyNi, (Tb$_{1/3}$Dy$_{1/3}$Ho$_{1/3}$)Ni, and (Gd$_{1/5}$Tb$_{1/5}$Dy$_{1/5}$Ho$_{1/5}$Er$_{1/5}$)Ni, which are predominantly composed of the FeB-type structure components.

In this paper, we report on the structural and magnetic properties of RNi (R=Dy, Tb$_{1/3}$Dy$_{1/3}$Ho$_{1/3}$, and Gd$_{1/5}$Tb$_{1/5}$Dy$_{1/5}$Ho$_{1/5}$Er$_{1/5}$). 
Our findings confirm that ferromagnetic ordering is robust, and that $T_\mathrm{C}$ is relatively unaffected by the increase of configurational entropy at the rare-earth site.
(Gd$_{1/5}$Tb$_{1/5}$Dy$_{1/5}$Ho$_{1/5}$Er$_{1/5}$)Ni shows an additional magnetic anomaly below $T_\mathrm{C}$, which suggests a possible spin reorientation. 
We evaluated the configurational entropy dependence of the magnetocaloric effect, which is discussed along with the anisotropy of rare-earth magnetic moments.

\begin{figure}
\centering
\includegraphics[width=0.8\linewidth]{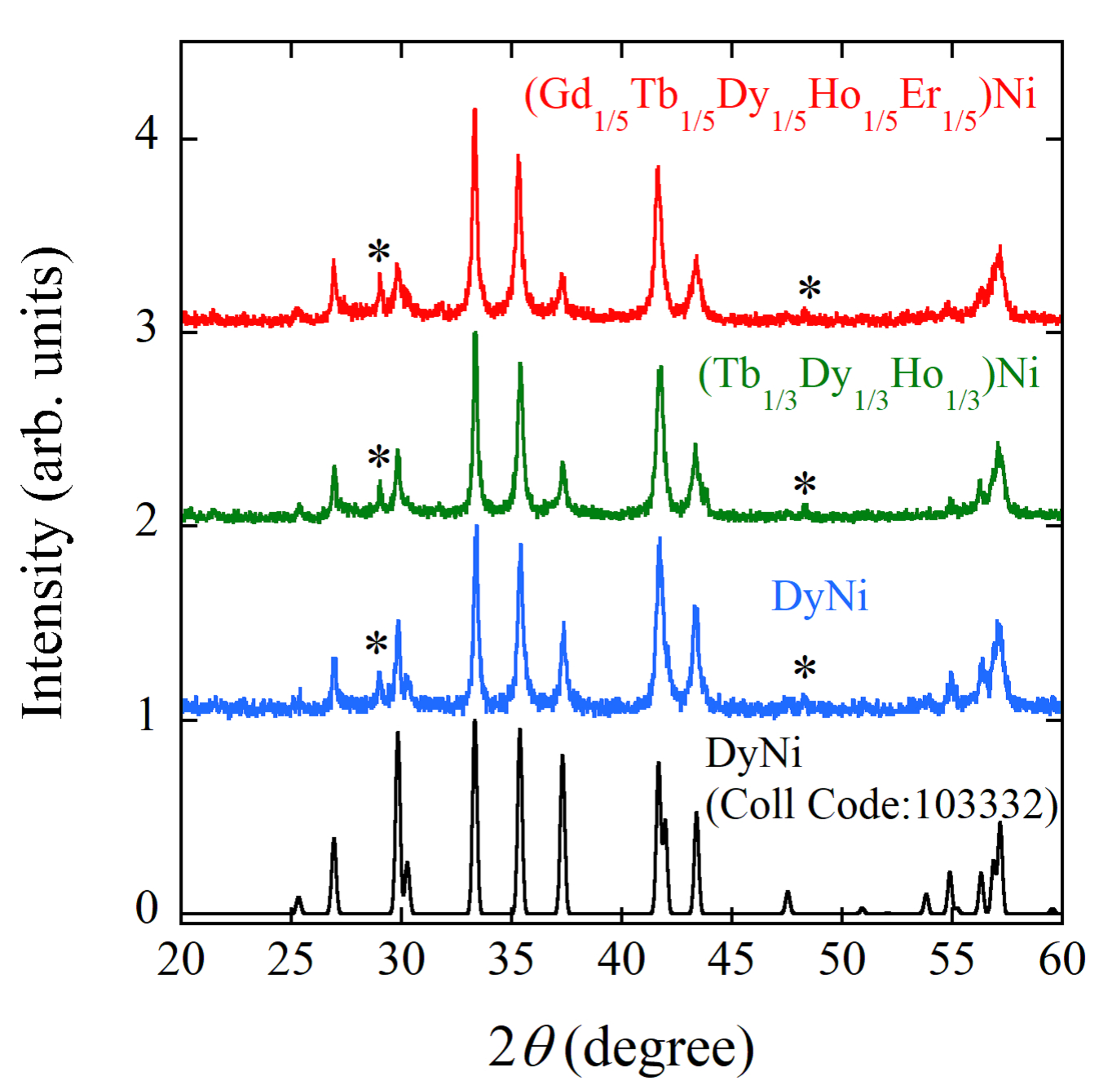}% Here is how to import EPS art
\caption{\label{fig1} XRD patterns of DyNi, (Tb$_{1/3}$Dy$_{1/3}$Ho$_{1/3}$)Ni, and (Gd$_{1/5}$Tb$_{1/5}$Dy$_{1/5}$Ho$_{1/5}$Er$_{1/5}$)Ni. The origin of each pattern is shifted by a value for clarity.}
\end{figure}

\section{Materials and Methods}
Polycrystalline samples were prepared using a homemade arc furnace as detailed in Table \ref{tab:table1}.
The materials used were rare earth (Gd, Tb, Dy, Ho, and Er) (99.9 \%) and Ni (99.9 \%).
The constituent elements with the stoichiometric ratio were melted on a water-cooled Cu hearth under an Ar atmosphere.
The button-shaped samples were remelted several times and flipped each time to ensure homogeneity.
Each as-cast sample was then annealed in an evacuated quartz tube at 800 $^{\circ}$C for four days.
Room temperature X-ray diffraction (XRD) patterns of powdered samples were obtained using an X-ray diffractometer (XRD-7000L, Shimadzu) with Cu-K$\alpha$ radiation.

The temperature dependence of dc magnetization $\chi_\mathrm{dc}$ ($T$) between 50 K and 300 K was measured using VSM (vibrating sample magnetometer) option of VersaLab (Quantum Design).
The isothermal magnetization curves between 50 K and 110 K were also taken using the VersaLab.

\section{Results and Discussion}
Figure \ref{fig1} displays the XRD patterns of DyNi, (Tb$_{1/3}$Dy$_{1/3}$Ho$_{1/3}$)Ni, and (Gd$_{1/5}$Tb$_{1/5}$Dy$_{1/5}$Ho$_{1/5}$Er$_{1/5}$)Ni, along with the simulation pattern of DyNi with the FeB-type structure taken from the ICSD database (Coll Code: 103332). 
All experimental patterns match the simulation pattern.
As mentioned in the Introduction, GdNi or TbNi crystallizes into a structure different from the FeB-type of RNi (R=Dy, Ho, and Er). 
However, the FeB-type structure is stabilized when dominant elements of Dy, Ho, and Er are present.
We note that the extra diffraction peaks assigned as the R$_{2}$O$_{3}$ (R=Dy, Tb+Dy+Ho, or Gd+Tb+Dy+Ho+Er) phase are detected (see * in Fig. \ref{fig1}).
Table \ref{tab:table1} lists the lattice parameters determined with the help of Rietveld refinement program\cite{Izumi:SSP2007,Tsubota:SR2017}.
While the $c$-axis length is almost independent of configurational entropy change at the rare-earth site, the $a$-axis (the $b$-axis) exhibits a slight expansion (contraction) with increasing configurational entropy.

\begin{figure}
\centering
\includegraphics[width=0.9\linewidth]{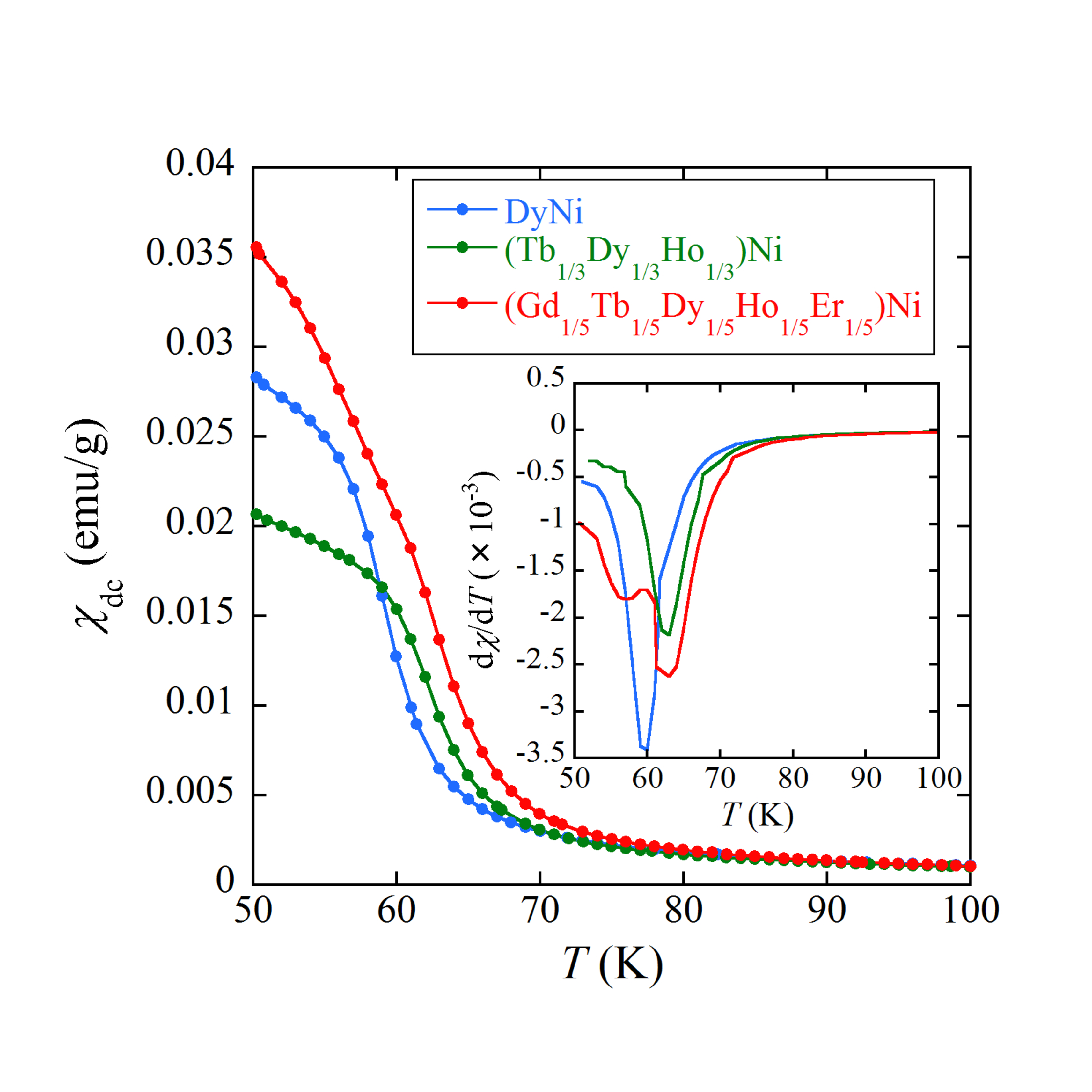}% Here is how to import EPS art
\caption{\label{fig2} Temperature dependences of $\chi_\mathrm{dc}$ of RNi system. The external field is 100 Oe. The inset is the temperature derivative of $\chi_\mathrm{dc}$ for each sample.}
\end{figure}

\begin{figure*}
\centering
\includegraphics[width=0.85\linewidth]{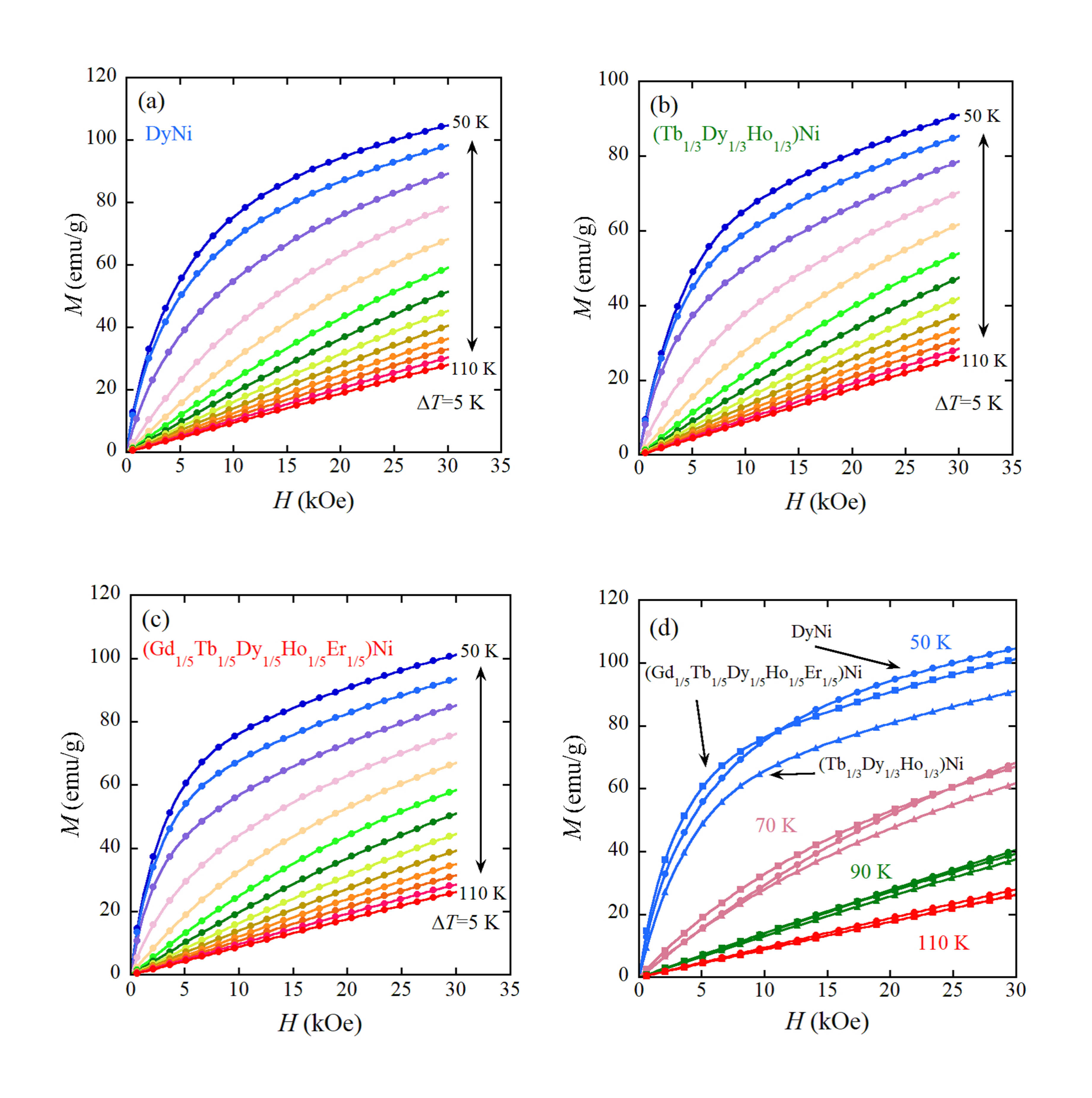}% Here is how to import EPS art
\caption{\label{fig3} Isothermal magnetization curves around $T_\mathrm{C}$ for (a) DyNi, (b) (Tb$_{1/3}$Dy$_{1/3}$Ho$_{1/3}$)Ni, and (c) (Gd$_{1/5}$Tb$_{1/5}$Dy$_{1/5}$Ho$_{1/5}$Er$_{1/5}$)Ni , respectively. (d) Comparison of magnetization curves among three RNi compounds at 50 K, 70 K, 90 K, and 110 K.}
\end{figure*}

Figure \ref{fig2} depicts $\chi_\mathrm{dc}$ ($T$) under an external field of 100 Oe for the RNi system.
Each sample exhibits a steep increase in $\chi_\mathrm{dc}$ below approximately 70 K, which is indicative of ferromagnetic ordering.
$T_\mathrm{C}$ is defined by the minimum point of the temperature derivative of $\chi_\mathrm{dc}$ (see the inset of Fig.\ref{fig2} and Table \ref{tab:table1}).
This is one of the effective ways to obtain $T_\mathrm{C}$ of ferromagnets\cite{Oikawa:APL2001,Miyahara:JSNM2018}.
DyNi undergoes a ferromagnetic transition at $T_\mathrm{C}$=59 K, which is consistent with the literature data\cite{Gignoux:JMMM1976}.
(Tb$_{1/3}$Dy$_{1/3}$Ho$_{1/3}$)Ni possesses $T_\mathrm{C}$=63 K, slightly enhanced compared to DyNi, and the $T_\mathrm{C}$ value remains unchanged in (Gd$_{1/5}$Tb$_{1/5}$Dy$_{1/5}$Ho$_{1/5}$Er$_{1/5}$)Ni.
We note that the $\chi_\mathrm{dc}$ ($T$) of (Gd$_{1/5}$Tb$_{1/5}$Dy$_{1/5}$Ho$_{1/5}$Er$_{1/5}$)Ni shows a small anomaly around 57 K, which is discussed later.
The results of $\chi_\mathrm{dc}$ ($T$) indicate that ferromagnetic ordering is resistant to atomic disorder at the rare-earth site.

DyNi, HoNi, and ErNi, which possess the orthorhombic FeB-type structure, exhibit a non-collinear  magnetic structure at $T_\mathrm{C}$ = 62 K, 37 K, and 13 K, respectively\cite{Rajivgandhi:JMMM2017,Gignoux:JMMM1976}. 
In these compounds, rare-earth magnetic moments have a ferromagnetic arrangement parallel to the $a$-axis and an antiferromagnetic arrangement parallel to the $c$-axis. 
The angle between the rare-earth moment and the $a$-axis is 29$^{\circ}$ for DyNi, 25$^{\circ}$ for HoNi, or 61$^{\circ}$ for ErNi\cite{Rajivgandhi:JMMM2017,Gignoux:JMMM1976}. 
Although the crystal structures of GdNi and TbNi differ from the FeB-type (GdNi: CrB-type, TbNi: monoclinic or orthorhombic)\cite{Rajivgandhi:JMMM2017}, they are also ferromagnets with $T_\mathrm{C}$ = 69 K and 67 K, respectively\cite{Rajivgandhi:JMMM2017}.
The magnetic ordering temperatures of RNi (=Dy, Ho, and Er) compounds follow the de Gennes scaling, which suggests a weak effect of energy-level splitting of the $J$-multiplet due to the crystalline-electric-field effect\cite{Bucher:PRB1975,Kitagawa:JALCOM1997} at $T_\mathrm{C}$.
In such a case, the 4$f$ electron distribution of a single rare-earth ion would be responsible for the magnetic structure\cite{Rinehart:ChemSci2011,Kitagawa:RP2019}. 
The 4$f$ electron distribution of a single R$^{3+}$ ion (R=Dy or Ho) is oblate, and the direction of the rare-earth magnetic moment is perpendicular to the equatorially expanded 4$f$-electron charge cloud\cite{Rinehart:ChemSci2011}.
On the other hand, the 4$f$ electron distribution of a single Er$^{3+}$ ion is prolate\cite{Rinehart:ChemSci2011}, causing the magnetic moment of Er ion to be perpendicular to that of the R$^{3+}$ ion (R=Dy or Ho).
In fact, the magnetic moments of DyNi and HoNi are nearly parallel to the $a$-axis and the direction of the Er$^{3+}$ moment tilts toward the $c$-axis.
The 4$f$ electron distribution of a single Tb$^{3+}$ ion is oblate, which is the same as Dy$^{3+}$ or Ho$^{3+}$.
Therefore, the magnetic structure of (Tb$_{1/3}$Dy$_{1/3}$Ho$_{1/3}$)Ni would not be significantly different from that of DyNi. 
However, in (Gd$_{1/5}$Tb$_{1/5}$Dy$_{1/5}$Ho$_{1/5}$Er$_{1/5}$)Ni, competition between easy magnetization axes might occur, potentially leading to a spin reorientation as observed in  (Gd$_{0.38}$Tb$_{0.27}$Dy$_{0.20}$Ho$_{0.15}$)Mn$_{6}$Sn$_{6}$\cite{Min:CP2022}.
As shown in Fig. \ref{fig2}, $\chi_\mathrm{dc}$ ($T$) of (Gd$_{1/5}$Tb$_{1/5}$Dy$_{1/5}$Ho$_{1/5}$Er$_{1/5}$)Ni shows a small anomaly around 57 K, which is clearly detected by d$\chi$/d$T$ with a double-dip structure (see the inset of Fig. \ref{fig2}).
We speculate that the anomaly at a lower temperature of 57 K suggests a change of magnetic structure like a spin reorientation.

The isothermal magnetization curves ($M$: magnetization and $H$: external field) measured around $T_\mathrm{C}$ are shown in Fig.\ref{fig3}(a) for DyNi, Fig. \ref{fig3}(b) for (Tb$_{1/3}$Dy$_{1/3}$Ho$_{1/3}$)Ni, and Fig.\ref{fig3}(c) for (Gd$_{1/5}$Tb$_{1/5}$Dy$_{1/5}$Ho$_{1/5}$Er$_{1/5}$)Ni, respectively.
In each sample, the pronounced steep increase of magnetization at lower external fields below approximately $T_\mathrm{C}$ supports the ferromagnetic ground state.
We note that the noticeable irreversibility is not observed in any of the samples.
Figure \ref{fig3}(d) provides a comparison of the magnetization curves among the three compounds at temperatures of 50 K, 70 K, 90 K, and 110 K.
With decreasing temperature, the $M$-$H$ curve of (Tb$_{1/3}$Dy$_{1/3}$Ho$_{1/3}$)Ni deviates from the other curves, albeit displaying a resemblance to that of (Gd$_{1/5}$Tb$_{1/5}$Dy$_{1/5}$Ho$_{1/5}$Er$_{1/5}$)Ni. 
As illustrated in Fig. \ref{fig2}, $\chi_\mathrm{dc}$ ($T$) of (Tb$_{1/3}$Dy$_{1/3}$Ho$_{1/3}$)Ni is smaller compared to the other two compounds at low temperatures, indicating a relatively weaker magnetic response in (Tb$_{1/3}$Dy$_{1/3}$Ho$_{1/3}$)Ni. 
Consequently, this might lead to the lowest $M$ for (Tb$_{1/3}$Dy$_{1/3}$Ho$_{1/3}$)Ni at a fixed $T$ (temperature) and $H$.
It should be noted that the variation in magnetic moment associated with each sample is another contributing factor to the differences in $M$ at fixed $T$ and $H$. 
Further investigation is required to elucidate the individual element's specific contribution.
Moreover, Fig. \ref{fig3}(d) reveals the intersection of magnetization curves between DyNi and (Gd$_{1/5}$Tb$_{1/5}$Dy$_{1/5}$Ho$_{1/5}$Er$_{1/5}$)Ni at 50 K or 70 K.
Such phenomena may be attributed to changes in magnetic anisotropy energy and saturation magnetic moment.

\begin{figure}
\centering
\includegraphics[width=0.8\linewidth]{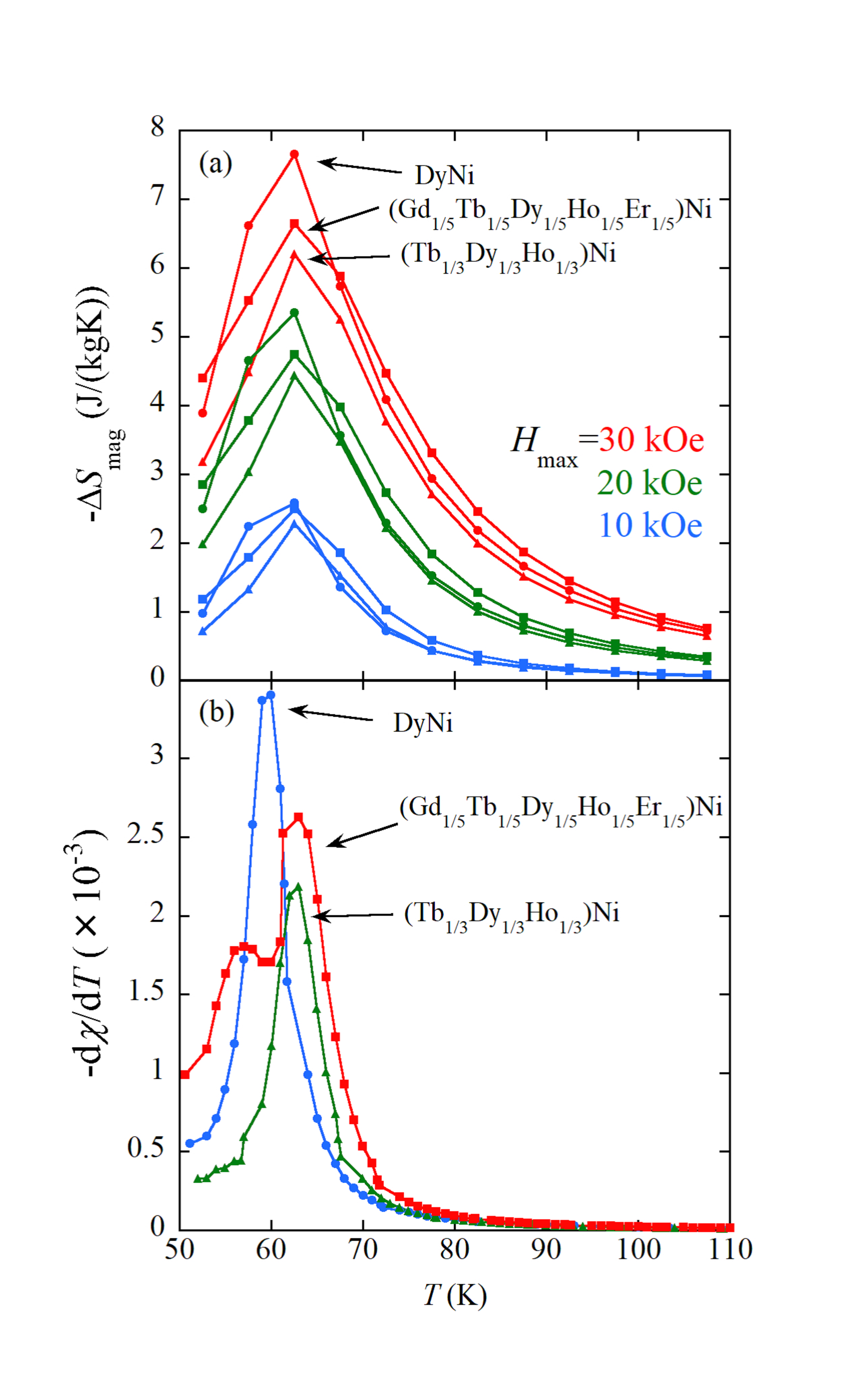}% Here is how to import EPS art
\caption{\label{fig4} (a) Temperature dependences of -$\Delta S_\mathrm{mag}$ at $H_\mathrm{max}$=10 kOe, 20 kOe, and 30 kOe for RNi system. (b) Temperature dependences of $-\mathrm{d} \chi_\mathrm{dc}/\mathrm{d} T$ for RNi system.}
\end{figure}

The magnetic entropy change $\Delta S_\mathrm{mag}$ ($T$,$H$) is obtained by using the Maxwell's relation as follows:
\begin{equation}
\Delta S_\mathrm{mag}(T,H)=\int_{0}^{H_\mathrm{max}}\left [\frac{\partial M(T,H)}{\partial T}\right ]_{H}dH
\label{equ:Smag}
\end{equation}
, where $H_\mathrm{max}$ is the maximum external field.
The temperature dependences of -$\Delta S_\mathrm{mag}$ ($T$) at $H_\mathrm{max}$=10 kOe, 20 kOe, and 30 kOe for the RNi system are summarized in Fig.\ref{fig4} (a).
All samples show a maximum of -$\Delta S_\mathrm{mag}$ ($T$) at approximately $T_\mathrm{C}$.
According to Eq. (1), $\Delta S_\mathrm{mag}$ ($T$) is influenced by $[\frac{\partial M(T,H)}{\partial T}]_{H}$.
This implies that a significant change in $M$ with decreasing temperature at a fixed $H$ is necessary to enhance $\Delta S_\mathrm{mag}$ ($T$).
Therefore, it is worthwhile to compare -$\Delta S_\mathrm{mag}$ ($T$) with $-\mathrm{d} \chi_\mathrm{dc}/\mathrm{d} T$ (refer to Figs.\ref{fig4} (a) and \ref{fig4} (b)), as the latter represents the change in the initial slope of the $M$-$H$ curve resulting from the temperature variations.
A larger value of $-\mathrm{d} \chi_\mathrm{dc}/\mathrm{d} T$ has the potential to contribute to a more significant change in $M$ when the temperature changes.
The dependence of $-\mathrm{d} \chi_\mathrm{dc}/\mathrm{d} T$ on configurational entropy resembles that of -$\Delta S_\mathrm{mag}$ ($T$), particularly at $H_\mathrm{max}$=10 kOe. 
At each $H_\mathrm{max}$, while the peak value of -$\Delta S_\mathrm{mag}$ for (Gd$_{1/5}$Tb$_{1/5}$Dy$_{1/5}$Ho$_{1/5}$Er$_{1/5}$)Ni diminishes compared to DyNi, the temperature dependence of -$\Delta S_\mathrm{mag}$ becomes broader. 
This broadening is advantageous for magnetic refrigeration applications. 
The presence of a spin reorientation below $T_\mathrm{C}$ contributes to this advantage, as the modification in magnetic structure gives rise to an additional $-\mathrm{d} \chi_\mathrm{dc}/\mathrm{d} T$.
As mentioned earlier, this spin reorientation likely arises from the interaction between rare-earth elements with distinct magnetic anisotropy. 
Consequently, the present study suggests the potential enhancement of magnetocaloric properties by manipulating rare-earth magnetic moment anisotropy in the high-entropy state.

In this discussion, we aim to compare the magnetocaloric effect between RNi and rare-earth HEAs. 
Specifically, we examine the peak value of -$\Delta S_\mathrm{mag}$, denoted as -$\Delta S^\mathrm{peak}_\mathrm{mag}$. 
In the RNi system, the configurational entropy dependence of -$\Delta S^\mathrm{peak}_\mathrm{mag}$ exhibits a non-systematic trend.
-$\Delta S^\mathrm{peak}_\mathrm{mag}$ decreases on going from DyNi, (Gd$_{1/5}$Tb$_{1/5}$Dy$_{1/5}$Ho$_{1/5}$Er$_{1/5}$)Ni to (Tb$_{1/3}$Dy$_{1/3}$Ho$_{1/3}$)Ni. 
In certain rare-earth HEAs\cite{Law:JMR2023}, -$\Delta S^\mathrm{peak}_\mathrm{mag}$ decreases in the order of GdTbDy, GdTbDyHo, GdTb, and GdTbDyHoEr (GdTbHoEr). 
In these rare-earth HEAs, changes occur in the average number of 4$f$ electrons and lattice constants, resulting in varying magnetic ordering temperatures ranging from 184 K to 258 K\cite{Law:JMR2023}. 
In contrast, our RNi system maintains a nearly constant $T_\mathrm{C}$, likely due to minimal alterations in lattice parameters and the average number of 4$f$ electrons. 
However, both RNi and rare-earth HEAs exhibit a non-systematic configurational entropy dependence of -$\Delta S^\mathrm{peak}_\mathrm{mag}$. 
Therefore, it appears that factors other than configurational entropy may influence the control of -$\Delta S^\mathrm{peak}_\mathrm{mag}$. 
Here we comment on the -$\Delta S^\mathrm{peak}_\mathrm{mag}$ value of (Gd$_{1/5}$Tb$_{1/5}$Dy$_{1/5}$Ho$_{1/5}$Er$_{1/5}$)Ni.
It is widely acknowledged that -$\Delta S^\mathrm{peak}_\mathrm{mag}$ follows a power law dependence on the magnetic field\cite{Franco:IJR2010}, represented as -$\Delta S^\mathrm{peak}_\mathrm{mag}$$\propto$$H^{n}$.
By applying this relation to (Gd$_{1/5}$Tb$_{1/5}$Dy$_{1/5}$Ho$_{1/5}$Er$_{1/5}$)Ni (see also Fig.\ref{fig4} (a)) and deducing the exponent $n$ to be 0.89, we can estimate -$\Delta S^\mathrm{peak}_\mathrm{mag}$ value at $H_\mathrm{max}$=50 kOe to be 10.6 J/kg$\cdot$K.
This value is larger compared to equimolar quinary rare-earth HEAs such as GdTbDyHoEr and GdTbHoErPr, which exhibit -$\Delta S^\mathrm{peak}_\mathrm{mag}$ values of 8.6 J/kg$\cdot$K and 6.92 J/kg$\cdot$K, respectively, at $H_\mathrm{max}$=50 kOe\cite{Yuan:AM2017,Lu:JALCOM2021}.

\section{Summary}
We have studied the effect of configurational entropy on the structural and magnetic properties of DyNi by successively replacing Dy with pair of R elements located on both sides of Dy in the periodic table.
This elemental substitution of Dy preserves the lattice parameters and average number of 4$f$ electrons.
Although the crystal structures of GdNi and TbNi differ from the FeB-type of RNi (R=Dy, Ho, and Er), all RNi (R=Dy, Tb$_{1/3}$Dy$_{1/3}$Ho$_{1/3}$, and Gd$_{1/5}$Tb$_{1/5}$Dy$_{1/5}$Ho$_{1/5}$Er$_{1/5}$) samples crystallize into the FeB-type structure.
$T_\mathrm{C}$ of DyNi is almost unchanged by increasing the configurational entropy at the rare-earth site, and the ferromagnetic ordering is robust under the high-entropy state.
In (Gd$_{1/5}$Tb$_{1/5}$Dy$_{1/5}$Ho$_{1/5}$Er$_{1/5}$)Ni, the additional magnetic anomaly is observed, which would be attributed to a spin reorientation resulting from the introduction of Gd+Er and the emergence of competing magnetic interactions.
The competition does not disrupt the ferromagnetic ordering, even in the high-entropy state, but rather leads to a spin reorientation transition.  
Furthermore, we assessed the magnetocaloric effect of the RNi system. 
Although the peak value of -$\Delta S_\mathrm{mag}$ of (Gd$_{1/5}$Tb$_{1/5}$Dy$_{1/5}$Ho$_{1/5}$Er$_{1/5}$)Ni is reduced compared to DyNi, the temperature dependence of -$\Delta S_\mathrm{mag}$ becomes broader. 
Additionally, we observed a strong correlation between the configurational entropy dependence of -$\Delta S_\mathrm{mag}$ ($T$) and that of -$\mathrm{d} \chi_\mathrm{dc}/\mathrm{d} T$. 
Hence, the broadening of -$\Delta S_\mathrm{mag}$ ($T$) in (Gd$_{1/5}$Tb$_{1/5}$Dy$_{1/5}$Ho$_{1/5}$Er$_{1/5}$)Ni can be attributed to the spin reorientation arising from the mixing of rare-earth elements with distinct magnetic anisotropy. 
Consequently, our study suggests the potential for enhancing the magnetocaloric properties by designing the anisotropy of rare-earth magnetic moments in the high-entropy state.

\section*{ACKNOWLEDGMENTS}
J.K. is grateful for the support provided by the Comprehensive Research Organization of Fukuoka Institute of Technology.

\section*{AUTHOR DECLARATIONS}
\subsection*{Conflict of Interest}
The authors have no conflicts to disclose.

\subsection*{Author Contributions}
Yuito Nakamura: Investigation, Formal analysis. Koshin Takeshita: Investigation, Formal analysis. Terukazu Nishizaki: Investigation, Formal analysis, Writing - reviewing \& editing. Jiro Kitagawa: Supervision, Formal analysis, Writing - original draft, Writing - reviewing \& editing. 

\subsection*{DATA AVAILABILITY}
The data that support the findings of this study are available
from the corresponding author upon reasonable request.

\end{document}